\documentclass[prb,preprint]{revtex4-1} 
\usepackage{amsmath}  
\usepackage{amsfonts} 
\usepackage{graphicx} 
\begin{document}

\title{Maximum range of a projectile thrown from constant-speed circular motion}
\author{Nikola Poljak}
\affiliation{Department of Physics, University of Zagreb, Croatia}
\date{\today}
\begin{abstract}
The problem of determining the angle at which a point mass launched from ground level with a given speed is a standard exercise in mechanics. Similar, yet conceptually and calculationally more difficult problems have been suggested to improve student proficiency in projectile motion. The problem of determining the maximum distance of a rock thrown from a rotating arm motion is presented and analyzed in detail in this text. The calculational results confirm several conceptually derived conclusions regarding the initial throw position and provide some details on the angles and the way of throwing (underhand or overhand) which produce the maximum throw distance.\\\\
To be published in \bf{The Physics Teacher}.
\end{abstract}

\maketitle 

\section{Introduction} 

The problem of determining the angle $\theta$ at which a point mass launched from ground level with a given speed $v_0$ will reach a maximum distance is a standard exercise in mechanics. There are many possible ways of solving this problem \cite{Tho84, Pal82, Por77}, leading to the well-known answer of $\theta = \pi/4$, producing a maximum range of $D_{\textrm{max}}=v_0^2/g$, with $g$ being the free-fall acceleration. Conceptually and calculationally more difficult problems have been suggested to improve student proficiency in projectile motion \cite{HarPW, Bro92}, with the most famous example being the Tarzan swing problem \cite{Bit05, Rav13}. The problem of determining the maximum distance of a point mass thrown from constant-speed circular motion is presented and analyzed in detail in this text. The calculational results confirm several conceptually derived conclusions regarding the initial throw position and provide some details on the angles and the way of throwing (underhand or overhand) which produce the maximum throw distance.

\begin{figure}[h!]
  \centering
    \includegraphics[width=0.6\textwidth]{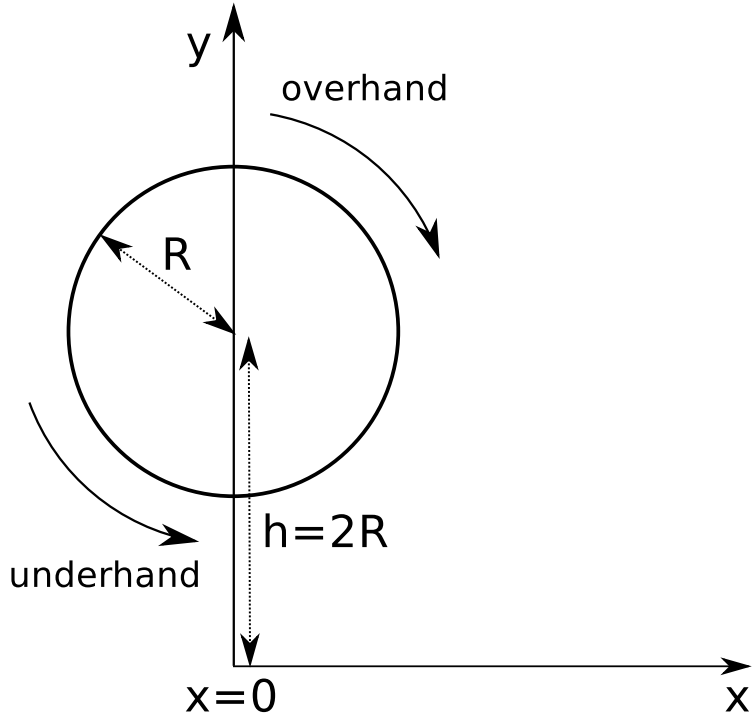}
\caption{An illustration of the stone throw problem.}
\label{figure1}
\end{figure}

The situation analyzed in this text can be defined as follows:{ \it``Suppose you want to throw a stone (approximated by a point mass) as far horizontally as possible. The stone rotates in a vertical circle with constant speed $v_0$. At which point during the motion should the stone be released? Should it rotate clockwise (an overhand throw) or counter-clockwise (an underhand throw)? The center of rotation is at height $h=2R$, where $R$ is the radius of rotation. The horizontal distance is measured from the point on the ground directly below the center of rotation.''} An illustration of the problem is given in Fig.\ref{figure1}.

This problem poses several conceptual difficulties. During motion, the initial height, the horizontal distance with respect to the reference point on the ground and the launch angle all change. Since all of these influence the final horizontal distance, it is not easy to deduce exactly what kind of throw should be executed to attain the maximum distance for a given speed. Let's assume that the throw is executed to the right (this does not restrict the solution in any way). For an overhand throw, the stone needs to be released during movement through the upper part of the circle, since then it is traveling to the right. During the first part of the motion, the angle of the stone velocity with the horizon is positive, so the stone is thrown upwards, but the initial horizontal distance from the reference point on the ground is negative. During the second part of the motion, the opposite holds true. It is clear that if the initial speed is nearly zero, the stone should be released when it is as forward as possible, since then it is practically released from rest and it drops vertically and lands at a distance $R$ away from the reference point. On the other hand, if the initial speed of the stone is very large, in the sense that the initial displacement from the reference point on the ground is very small compared to the range of the throw, one would expect that the classical result of an angle equal to $\theta = \pi/4$  produces the largest horizontal distance. 

For an underhand throw, the the stone needs to be released during movement through the lower part of the circle, since then it is traveling to the right. In this case, it is more obvious that the release should happen during the second part of the motion, since then the throw is executed upwards and the initial horizontal displacement of the stone is positive. Again, for a low speed, the stone should be released when it is as forward as possible and for a high speed, it should be released when the throw angle with respect to the horizon is equal to $\theta = \pi/4$. Interestingly, it is unclear whether the throw should be made overhand or underhand to obtain the maximum throw distance for a general value of speed. To answer this question, some knowledge of elementary kinematics and numerical calculation is necessary.

\section{Methods}

\begin{figure}[h!]
  \centering
    \includegraphics[width=0.7\textwidth]{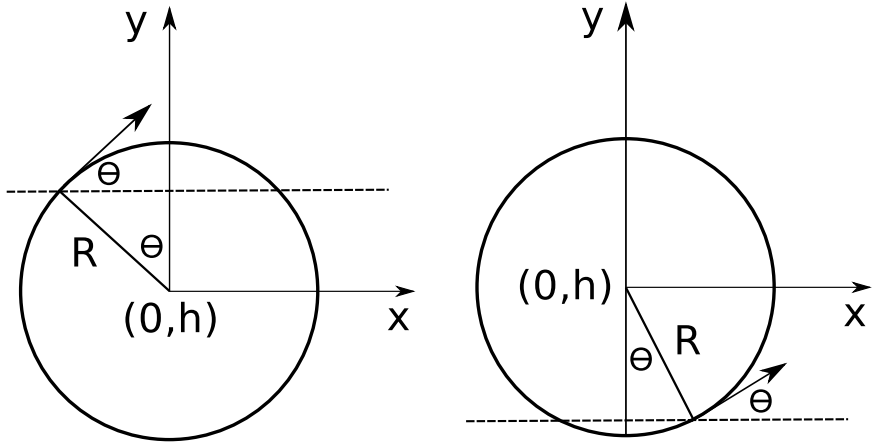}
\caption{The overhand (left) and the underhand (right) throws in the coordinate system. $\theta$ marks the angle the stone velocity makes with the horizontal line.}
\label{figure2}
\end{figure}

We can define the coordinate system as in Fig.\ref{figure2}. Clearly, there are two cases to consider: the overhand and the underhand throw. 

Let $\theta$ mark the angle with the horizontal line when the stone is released (set equal to $t=0$). The initial coordinates of the stone are:
\begin{eqnarray}
\label{cases}
(x_0, y_0)=\begin{cases} (-R\sin\theta, h+R\cos\theta) & \textrm{for an overhand throw, and} \\
(R\sin\theta, h-R\cos\theta) & \textrm{for an underhand throw.}\end{cases}
\end{eqnarray}

The moment when the stone hits the ground is found by setting $y(t_{\textrm{r}})=0$ in the general equation of motion $y=y_0+v_0t\sin\theta-\frac{1}{2}gt^2$, which gives one physical solution for $t_{\textrm{r}}$. Inserting that solution into $x=x_0+v_0t\cos\theta$, the throw horizontal distance becomes:
\begin{eqnarray}
&D=x_0+\frac{v_0^2\cos\theta}{g}\left(\sin\theta + \left|\sin\theta\right|\sqrt{1+\frac{2gy_0}{v_0^2\sin^2\theta}}\right)\cr
&= \mp R\sin\theta+\frac{v_0^2\cos\theta}{g}\left(\sin\theta + \left|\sin\theta\right|\sqrt{1+\frac{2g(h\pm R\cos\theta)}{v_0^2\sin^2\theta}}\right) \,.
\label{eq2}
\end{eqnarray}

The absolute sign is required to take into account possible negative values of angle $\theta$. The notation in which the upper sign represents the overhand throw and the lower represents the underhand throw is introduced. The trajectory equations here assume no air drag.

The maximum distance of the throw can be found numerically or graphically by plotting $D$ as a function of the inital speed $v_0$ and throw angle $\theta$. Another approach is to set $\textrm{d}D/\textrm{d}\theta=0$ for a certain intial speed $v_0$, which often has the benefit of emphasizing the dimensionless variables relevant to the problem. Taking the derivative one obtains the following condition, after some algebra:
\begin{eqnarray}
\label{eq_condition}
\cot\theta_{\textrm{m}}\left(A\cos\theta_{\textrm{m}} \mp 1\right)=A\sin\theta_{\textrm{m}}+\sqrt{(A\sin\theta_{\textrm{m}})^2+2A(B\pm \cos\theta_{\textrm{m}})}\cr
\hskip 35mm -\frac{A\cos\theta_{\textrm{m}}(A\cos\theta_{\textrm{m}} \mp 1)}{\sqrt{(A\sin\theta_{\textrm{m}})^2+2A(B\pm \cos\theta_{\textrm{m}})}}\,,
\end{eqnarray}
where the shorthands $A=v_0^2/(gR)$ and $B=h/R$ were introduced and $\theta_{\textrm{m}}$ denotes the throw angle at which the range is maximal. At this point, we will use the simplification $B=h/R=2$ and note that in that case $A$ is twice the ratio of the kinetic energy of the stone to its gravitational potential energy at the lowest point of rotation in the chosen coordinate system. Even though numerical solving was skipped in (\ref{eq2}), here it needs to be employed. The maximum angle results are obtained when varying the value of $A$ from 0 to 50, separately for the overhand and the underhand cases. The value of $A=0$ corresponds to the case where the stone is released from rest, while $A=50$ corresponds to the case in which the kinetic energy of the stone is much larger than its initial potential energy. At the end of the calculation, the throw distance can be found for a specific $\theta_{\textrm{m}}(A)$ angle value.

\section{Results}

\begin{figure}[h!]
  \centering
    \includegraphics[width=0.45\textwidth]{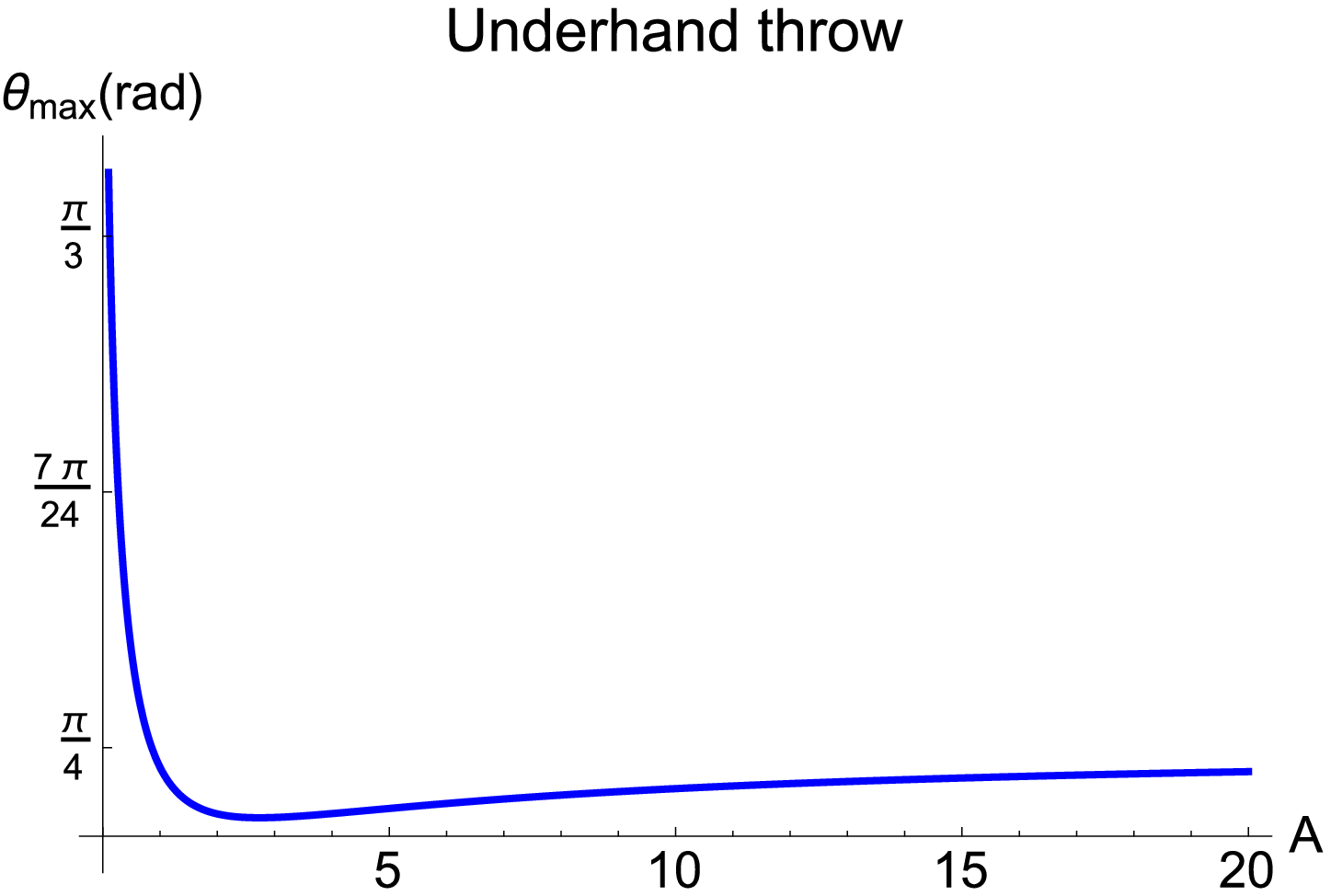}
    \includegraphics[width=0.45\textwidth]{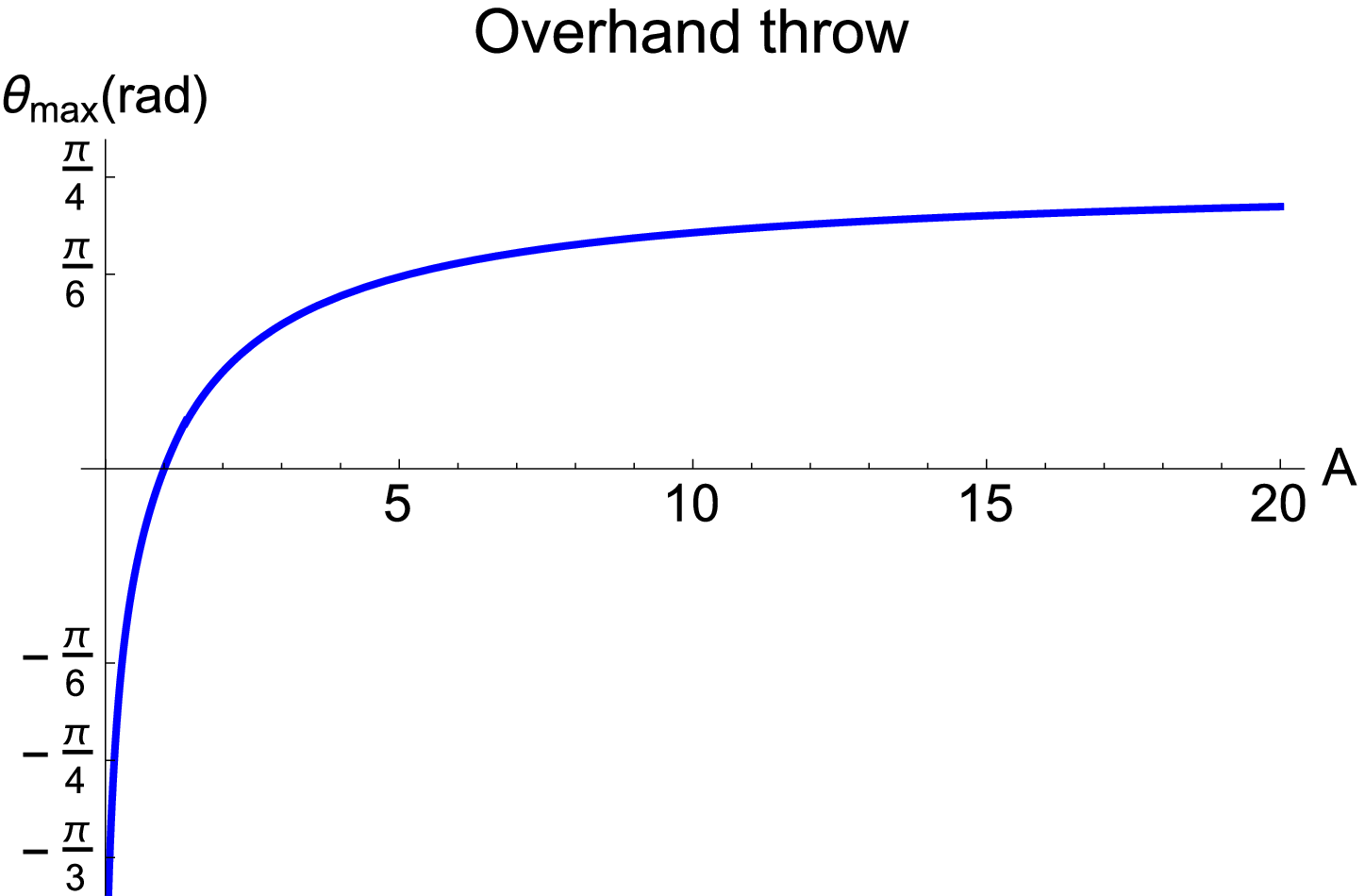}
\caption{The angle $\theta_{\textrm{max}}$ at which the throw distance is maximum for a given initial speed in cases of an underhand and an overhand throw. The parameter $A=v_0^2/{gR}$, where $v_0$ is the initial speed. The range of $A$ in the graphs is from 0.03 to 20.}
\label{figure3}
\end{figure}

The results of numerically solving (\ref{eq_condition}) are given on Fig.\ref{figure3}. As suspected earlier, the angles obtained for an underhand throw approach $\theta_{\textrm{m}} = \pi/2$ for small values of $A$ and $\theta_{\textrm{m}} = \pi/4$ for large values of $A$. Interestingly, the decline from $\pi/2$ to $\pi/4$ is not uniform. With $A$ increasing from zero, the angle of maximum distance decreases until $A$ reaches a minimum value at $A\approx 2.73$, corresponding to a minimum angle of $\theta_{\textrm{m}} = 0.749$, and then starts increasing again until it asymptotically reaches $\pi/4$. 

For an overhand throw, the maximum distance for very small values of $A$ is obtained when $\theta_{\textrm{m}} \approx -\pi/2$, as concluded earlier. With increasing $A$, the throw angle decreases to zero. A throw at $\theta_{\textrm{m}} = 0$, which happens at the top of the motion, is achieved when $A=1$, i.e. when $v_0=\sqrt{gR}$. Just as in the case of an underhand throw, when further increasing $A$, the throw velocity angle approaches $\theta_{\textrm{m}} = \pi/4$.

\begin{figure}[h!]
  \centering
    \includegraphics[width=0.45\textwidth]{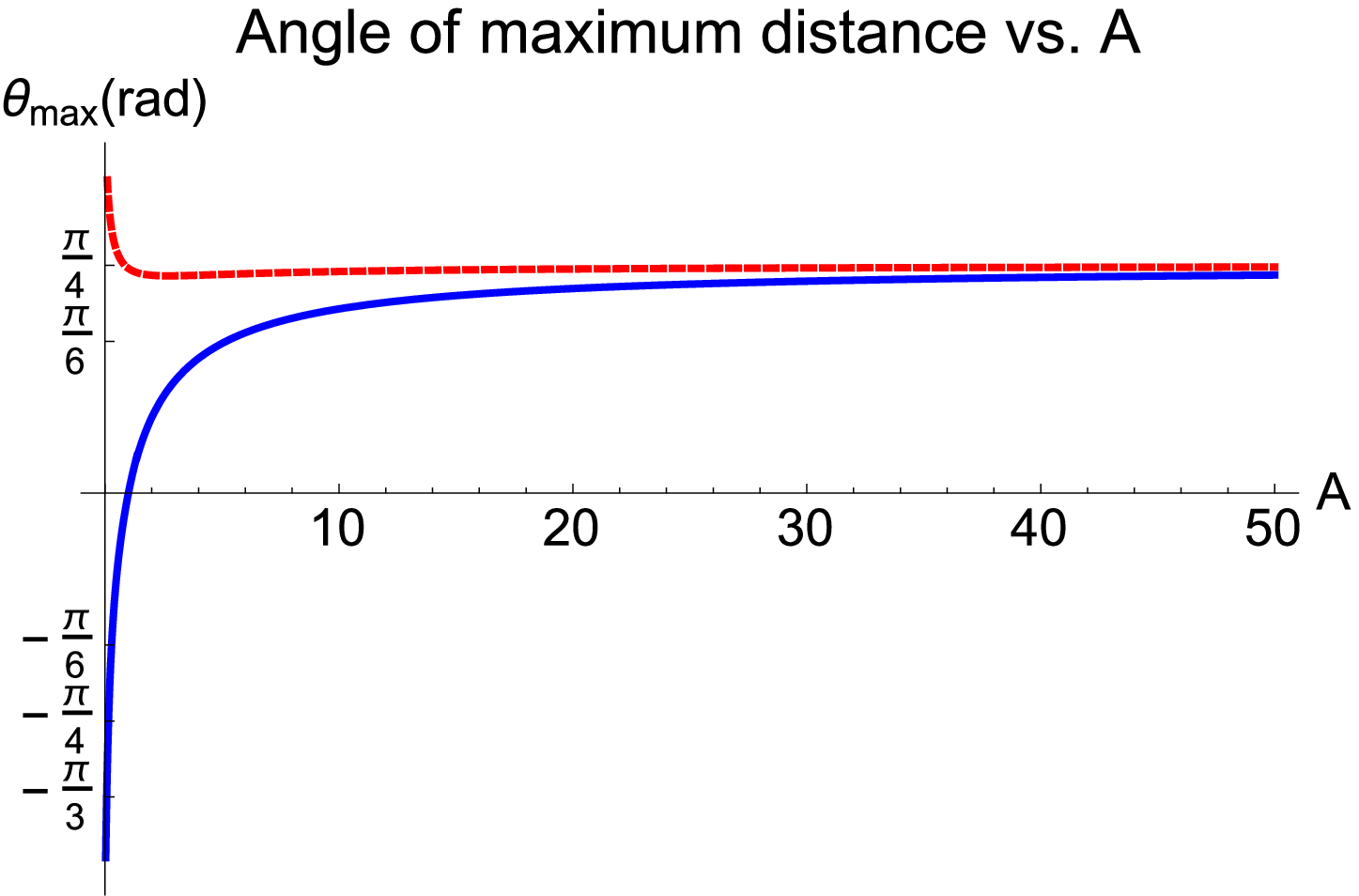}
    \includegraphics[width=0.45\textwidth]{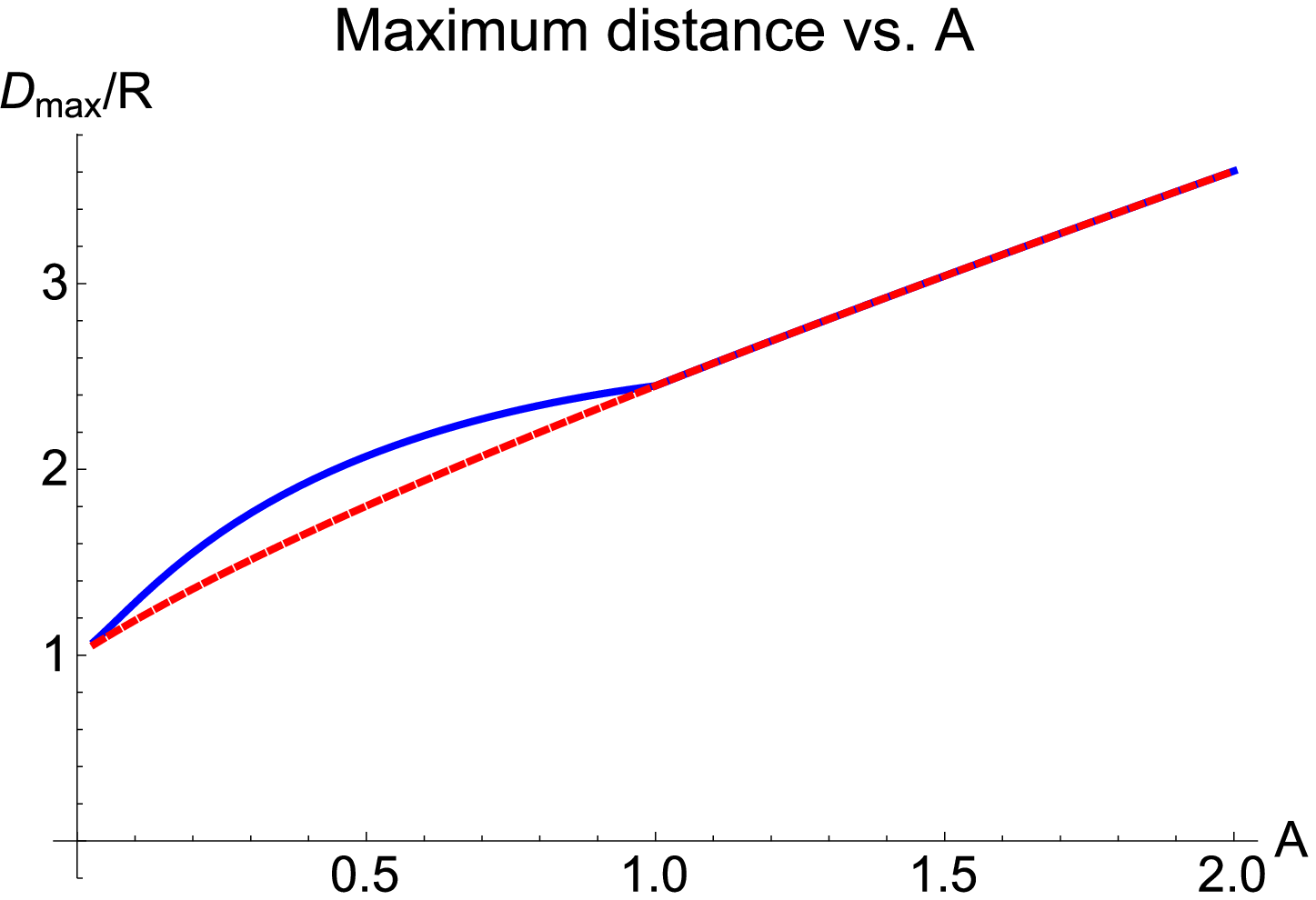}
\caption{(color online) Left: The angle $\theta_{\textrm{max}}$ at which the throw distance is maximum for a given initial speed. Right: The scaled maximum horizontal distance $D_{\textrm{max}}/R$ covered by the stone when launched with a given initial speed. The dashed red line represents an underhand throw, the solid blue line an overhand throw. On the right graph, $A$ ranges from 0.03 to 2.}
\label{figure4}
\end{figure}

Using the values for the obtained angles, one can calculate the maximum throw distances, as shown on Fig.\ref{figure4}. It is interesting to note that for speeds for which $A \geq 1$ and $A=0$, the maximum distances are the same in the cases of an overhand and an underhand throw. In the interval $0 < A < 1$, the maximum attainable distance is greater in the case of an overhand throw. This result could not be simply deduced without calculation, since after an overhand throw for which $A<1$ the stone has an initial velocity pointing downwards, decreasing its fall time and thus its final horizontal distance.

\section{Discussion}

The results of the calculation show several interesting features. For a throw speed equal to zero, the stone should be held as far forward as possible before release, performing a simple vertical drop. The distance it reaches in this case is equal to $R$. When increasing the speed in the interval $0 < v_0 < \sqrt{gR}$, it is more beneficial to perform an overhand throw, even though the stone release is performed when its velocity is pointing downwards. For $v_0 \geq \sqrt{gR}$, the maximum distances for an overhand and an underhand throw become exactly the same, but the throws are performed at different angles. For example, in the limiting case of $v_0=\sqrt{gR}$, an overhand throw is performed at the top of the motion (when $\theta_{\textrm{m}}=0$), while an underhand throw is performed when $\theta_{\textrm{m}} \approx 0.775$. With increasing speed, the maximum distance increases almost linearly with $A$ and the angles for both types of throws tend towards $\pi/4$. 

For an underhand throw, the angle at which the maximum distance is attained dips below its asymptotic value to the value of $\theta_{\textrm{m}} = 0.749$ before it starts increasing towards $\pi/4$. This result might present a conceptual difficulty -- after all, for a large enough speed, is it not always more beneficial that the throw is performed at $\theta=\pi/4$, since in that case the stone is both more forward and higher? Even though this statement is true, one should remember that the stone is not thrown from the ground level, and so the angle at which it attains maximal distance can not be exactly $\pi/4$. However, it is worth noting that in the case of an underhand throw, the convergence to $\pi/4$ seems very rapid (when compared to the overhand throw). 

\begin{figure}[h!t!]
  \centering
    \includegraphics[width=0.7\textwidth]{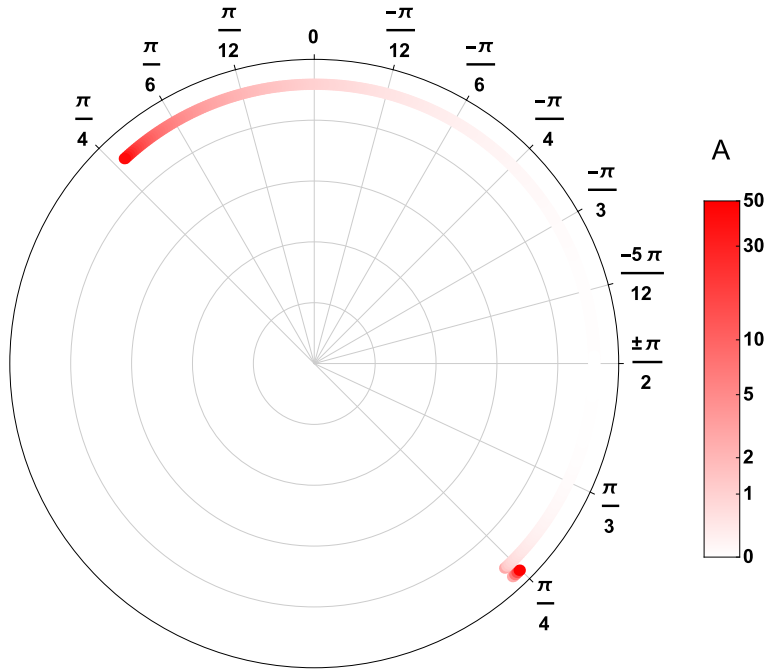}
\caption{(color online) The correspondence of the maximum distance angle and $A$ for overhand throws (top part) and underhand throws (bottom part of the circle).}
\label{figure5}
\end{figure}

All the results are summarized on Fig.\ref{figure5}. The circumference of the circle shows the points through which the stone can move. The angles given are the angles which the stone has with respect to the horizontal line when released from a certain point. The top part of the circle represents overhand throws and the bottom part underhand throws. Using the color scale, for a given value of $A$ one can find two corresponding angles on the circle which produce the maximum throw distance. The color scale is logarithmic since large arcs of the circle correspond to small changes in $A$. It can be seen that for $A=0$ the stone is released when it is furthest to the right. With increasing $A$, the angle grows for an overhead throw, asymptotically reaching $\pi/4$. For underhand throws, the angle first decreases below $\pi/4$ and then increases back towards it, as discussed. The decrease and the increase are drawn separately on the graphic to avoid overlap.

\section{Conclusion}

The problem of throwing a point particle from a vertical circular constant-speed motion was considered. The optimal throw angle $\theta_{\textrm{m}}$ which maximizes the throw distance was found as a function of $A=v_0^2/(gR)$ and of the throw type. The results confirmed some conceptually derived predictions, but also provided some new insights. To obtain the maximum throw distance when $A<1$ it is more beneficial to use the overhand throw. Interestingly, for $A \geq 1$ the maximum throw distance becomes the same for an overhand and an underhand throw, although the throws are executed at different angles with respect to horizon. For very large values of $A$, the optimal angle in both cases equals $\pi/4$ and the throw distance becomes $D_{\textrm{max}} \approx v_0^2/g$. The result is approximate since the throw is not performed from ground level.

If one wants to throw a stone as far as possible in real life, many effects such as air drag and the spin of the stone, among others, need to be considered. Furthermore, vertical circular motion is not a good model for human overhand throwing which is a complex sequence of torques and levers that can generate more initial speed than underhand throwing. Nevertheless, it would be interesting to construct an apparatus to experimentally demonstrate the results outlined here, as well as evaluate the effects due to air drag and other factors, which is considered as a future step of this article.

\end{document}